\documentclass{iaus}
\usepackage{graphicx}

\title[Time variation of radial abundance gradients] 
{New results on the time variation of the radial abundance gradients
from planetary nebulae 
\thanks{Work partially supported by CNPq and FAPESP.}}

\author[Maciel, Lago and Costa]   
{Walter J. Maciel$^1$, L. G. Lago$^1$ \break \and R. D. D. Costa$^1$}

\affiliation{$^1$IAG/USP, S\~ao Paulo SP, Brazil\break email: maciel@astro.iag.usp.br\\}

\pubyear{2006}
\volume{234}  
\pagerange{1--2}
\date{?? and in revised form ??}
\jname{Planetary Nebulae in our Galaxy and Beyond}
\editors{M.J. Barlow \& R.H. M\'endez, eds.}
\begin{document}

\maketitle

\begin{abstract}
New results on the time variation of the radial abundance gradients 
in the galactic disk are presented on the basis of four different 
samples of planetary nebulae. These comprise both smaller, homogeneous 
sets of data, and larger but non-homogeneous samples. Four different 
chemical elements are considered, namely, O, S, Ar, and Ne. 
Other objects such as open clusters, cepheids and HII
regions are also taken into account. Our analysis support our earlier 
conclusions in the sense that, on the average, the radial abundance 
gradients have flattened out during the last 6 to 8 Gyr, with important
consequences for models of the chemical evolution of the Galaxy.
\keywords{planetary nebulae, chemical evolution, abundances}
\end{abstract}

\firstsection 
\section{Introduction}

The time variation of the radial abundance gradients is possibly the most
important information that can be obtained from the abundance  variations 
in the galactic disk. In this respect, planetary nebulae (PNe) play a 
particularly important role, as they have relatively well determined 
abundances  and are originated from stars within a reasonably 
large mass (and age) bracket (see for example Maciel \& Costa \cite{mc03}).

Using PNe, Maciel et al. (\cite{mcu2003}) suggested  
a time flattening of the O/H gradient from roughly $-$0.11 dex/kpc 
to $-$0.06 dex/kpc during the last 9~Gyr. More recently, Maciel et al. 
(\cite{mlc2005a}) extended the original discussion by  
estimating the  [Fe/H] gradient and using other objects such as open 
clusters, cepheid variables, HII regions and stars in OB associations.
In the present work, we consider four chemical elements  (O, S, Ar, and 
Ne) and take into account four different PNe samples. A detailed 
discussion can be found in Maciel et al. (\cite{mlc2006}).

\section{The Data}

The four samples considered are: (i) the basic sample, which is essentially 
the same used in our previous work (Maciel et al. \cite{mcu2003}, 
\cite{mlc2005a}, \cite{mlc2005b}, Costa et al. \cite{costa04}). 
This sample contains up to 234 objects and is the largest one considered, 
but it is a compilation, albeit  careful, of several different determinations 
in the literature; (ii) the homogeneous sample of Henry  et al. (\cite{henry04}); 
(iii) the sample recently presented by Perinotto et al. (\cite{perinotto}), 
and (iv) our own data, which we will call the IAG/USP sample. This is a 
highly homogeneous sample, although relatively small, reaching about 70 nebulae 
(see Costa et al. \cite{costa04} for details).

\section{Results and Discussion}

Abundance gradients were determined for O, S, Ar, and Ne assuming a linear 
variation in the abundances with the galactocentric distance $R$. 
The ages of the PNe progenitors were estimated in the following way. 
First, the heavy element abundances were converted into [Fe/H] 
metallicities and then the ages were determined using an age-metallicity 
relationship (AMR) which also depends on the galactocentric distance. 
Once the individual ages had been determined, the nebulae in each sample 
were divided into two age groups,  Group~I (younger) and  Group~II (older). 
We considered the age separation of the groups $t_I$ in the range 
3.0 to 6.0 Gyr, and for each of these values we calculated  the gradients 
of Groups I and II. Comparing the derived gradients for all four 
elements and samples, we conclude that the gradients of the younger 
Group~I are systematically flatter than the corresponding gradients 
of the older Group~II. Considering the other objecs studied in our 
previous work, namely open clusters, cepheid variables, HII regions 
and stars in OB associations, the general picture of the time variation 
of the gradients becomes more clear, as can be seen in Fig.~1.
Predictions of models by Hou et al. (\cite{hou}) and Chiappini 
et al. (\cite{cmr2001}) are also shown, as illustrations of theoretical 
models of chemical evolution. It can be concluded that the flattening 
rate is essentially the same as derived before, namely,  
$d[{\rm Fe/H}]/dR \sim 0.005$ to $-0.010\,$dex kpc$^{-1}$ Gyr$^{-1}$ for 
the last 6 to 8 Gyr.

\begin{figure}
\centering
\includegraphics[height = 3 in, angle=-90]{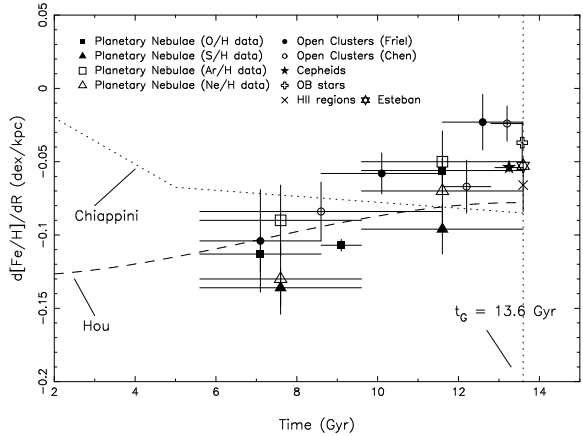}
  \caption{Time variation of the [Fe/H] gradient using PNe, 
   open clusters, HII regions, Cepheids and OB stars.
   Theoretical models by Hou et al. (2000) and Chiappini et
   al. (2001) are also shown.  The age of the galactic disk 
  is taken as 13.6 Gyr.}
\end{figure}

\end{document}